\documentclass[aps,prd,showpacs,nobibnotes,
nofootinbib,superscriptaddress]{revtex4}

\usepackage{graphicx}
\usepackage{amsmath}
\usepackage{latexsym}

\begin{document}

\title{Testing the Copernican Principle with Hubble Parameter}
\author{Zhi-Song Zhang}
\affiliation{Department of Aerospace Engineering, School of Astronautics, Harbin Institute of Technology (HIT), Harbin, Heilongjiang 150001, China}
\affiliation{Department of Astronomy, Beijing Normal University,
Beijing 100875, China}
\author{Tong-Jie Zhang}
\email{tjzhang@bnu.edu.cn}
\affiliation{Department of Astronomy, Beijing Normal University,
Beijing 100875, China}
\affiliation{Departments of Physics and Astronomy, University of California, Berkeley, CA 94720, USA}
\affiliation{Lawrence Berkeley National Laboratory, 1 Cyclotron Road, Berkeley, CA 94720, USA}
%\affiliation{Kavli Institute for Theoretical Physics China, CAS, Beijing
%100190, China}
\author{Hao Wang}
\affiliation{Department of Astronomy, Beijing Normal University,
Beijing 100875, China}
\author{Cong Ma}
\affiliation{Department of Astronomy, Beijing Normal University,
Beijing 100875, China}

\begin{abstract}
Using the longitudinal expression of Hubble expansion rate for the general Lema\^{i}tre-Tolman-Bondi (LTB)
metric as a function of cosmic time, we examine the scale on which the Copernican Principle
holds in the context of a void model. By way of performing parameter estimation on the CGBH void model,
we show that the Hubble parameter data favors a void with characteristic radius of
$2 \sim 3$ Gpc. This brings the void model closer, but not yet enough,
to harmony with observational indications given by the background kinetic Sunyaev-Zel'dovich
effect and the normalization of near-infrared galaxy luminosity function. However,
the test of such void models may ultimately lie in the future detection of
the discrepancy between longitudinal and transverse expansion rates, a touchstone of
inhomogeneous models. With the proliferation of observational Hubble parameter data
and future large-scale structure observation, a definitive test could be performed on
the question of cosmic homogeneity. Particularly, the spherical LTB void models
have been ruled out, but more general non-spherical inhomogeneities still need
to be tested by observation. In this paper, we utilise a spherical void model
to provide guidelines into how observational tests may be done with more general
models in the future.

\end{abstract}
\pacs{98.65.Dx, 98.80.Es, 98.62.Py}

\maketitle

\section{Introduction}
The Copernican Principle (CP) is the hypothesis that we do not occupy a
privileged position in the Universe.  It leads to the
Friedmann-Robertson-Walker (FRW) metric as the metric of the homogeneous and
isotropic background spacetime \cite{WaldRM1984}.  However, one may not expect
the CP to hold on all scales of cosmological interest, for both theory and
observation shows that large-scale structure can emerge even if a homogeneous
and isotropic initial background is assumed.  Recently, the observed
near-infrared luminosity function from a complete sample of galaxies indicates
that the data cannot rule out the possibility of our vicinity being described
by a void model \cite{2012arXiv1207.1588K}.  In addition, the void model may
also serve as a possible explanation to the emergence of accelerated expansion
of the Universe without employing an exotic component dubbed `dark energy'.  To
further investigate the ramifications of such a non-CP scenario and ascertain
the possible existence of a local void, we consider a variety of other
cosmological tests, as laid out in this paper.  We mainly make use of
the observational Hubble parameter data (OHD) which is independent of CMB and
galaxy distribution measurement, and their
observational properties have not been elucidated well enough in the
inhomogeneous void model.

Although the spherical symmetric LTB void models
have been ruled out, this does not imply that inhomogeneity, as a
whole, has "died"  as an alternative to the concordance model. Apparently, LTB voids
are simple and mathematically tractable, but its inherent spherical inhomogeneity
is very special and certainly non--generic. It is then
impossible to know {\it a priori} if the more generic forms and profiles of
inhomogeneity will fare like LTB voids and end up failing to fit the joint
current tests such as SN, the kSZ effect or BAO or OHD and so on. Thus, we in this paper still use spherical LTB models to provide a pathfinder of how future
non-spherical models could be tested in the future.

\section{LTB dynamics and the void model}
The Lema\^{i}tre-Tolman-Bondi (LTB) line element reads
\begin{equation}
    \label{Eq.ltbmetric}
    ds^2 = -dt^2 + \frac{A'(r, t)^2}{1 - k(r)}dr^2 + A^2(r, t)d\Omega^2,
\end{equation}
where $'$ denotes $\partial / \partial r$, and $k(r)$ is associated
with the spatial curvature. The Friedmann-Robertson-Walker (FRW) metric can be
recovered by imposing $A(r, t) = a(t) r$ and $k(r) = k r^2$.
Besides the most popular inhomogeneous exact solution of LTB model in cosmology,
another interesting family of that are
those found by Szekeres\cite{1975CMaPh..41...55S}, which
are much less idealised than spherical LTB models. Generally, these models have no symmetries (i.e. no killing-vectors \cite{1977GReGr...8..549B}) and are constructed by six arbitrary metric functions: one freedom being represented to rescale the ¡®radial¡¯ coordinate and remaining five degrees of freedom to model inhomogeneity. In fact, all of LTB quantities given in coordinate independent manner can be readily generalised to that in Szekeres models\cite{2012CQGra..29f5018S,2012JCAP...12..001W}.
The Gpc-size spherical symmetric LTB void modes are able to fit CMB data without dark energy under assumption that our cosmic observing position is very close to the void centre\cite{2011PhLB..697..265B,2013PhRvD..87b3524B}. This
certainly leads to an unacceptable fine tuning and is a direct effect of
spherical symmetry, and can also be corrected by considering non--spherical models. As shown in
\citep{2011PhLB..697..265B}, even the still idealised deviation from
spherical symmetry furnished by a quasi--spherical Szekeres model allows for a
significant improvement on this fine tuning of the centre position that has
always plagued LTB models.

From the LTB metric one can go on writing down and solving the dynamical
equations for LTB void models. One notices along the way that the spherical
symmetric configuration gives rise to two expansion rates
\begin{align} \label{Eq.Hperp} H_\bot \equiv \frac{\dot A}{A}, \quad
    H_\parallel \equiv
    \frac{\dot A'}{A'}
\end{align}

After choosing a gauge $A_0(r) = r$, and a homogeneous `bang time', one needs
only the boundary conditions to finally obtain the evolution history(see
Ref.~\cite{2000A&A...353...63C, 2012ApJ...748..111W} for more detailed
treatments). Expressed as two functions, $\Omega_m$ and $H_{\bot 0}$, these
boundary conditions define an LTB void model. Throughout this work, we employ
the Constrained GBH
(CGBH) model \cite{2008JCAP...04..003G}, in which
\begin{equation} \label{Eq.modelom}
    \Omega_m(r) = 1 + (\Omega_{0}-1) \left( \frac{1 - \tanh[(r - r_0) / 2
    \Delta{r}]}{1 + \tanh(r_0 / 2\Delta{r})} \right),
\end{equation}
where $\Omega_{0}$ describes the density at the symmetric center, $r_0$ is the
characteristic size of the void, and $\Delta r$ describes the steepness of the
void near the edge.

\begin{figure}
    \includegraphics[width=0.48\textwidth]{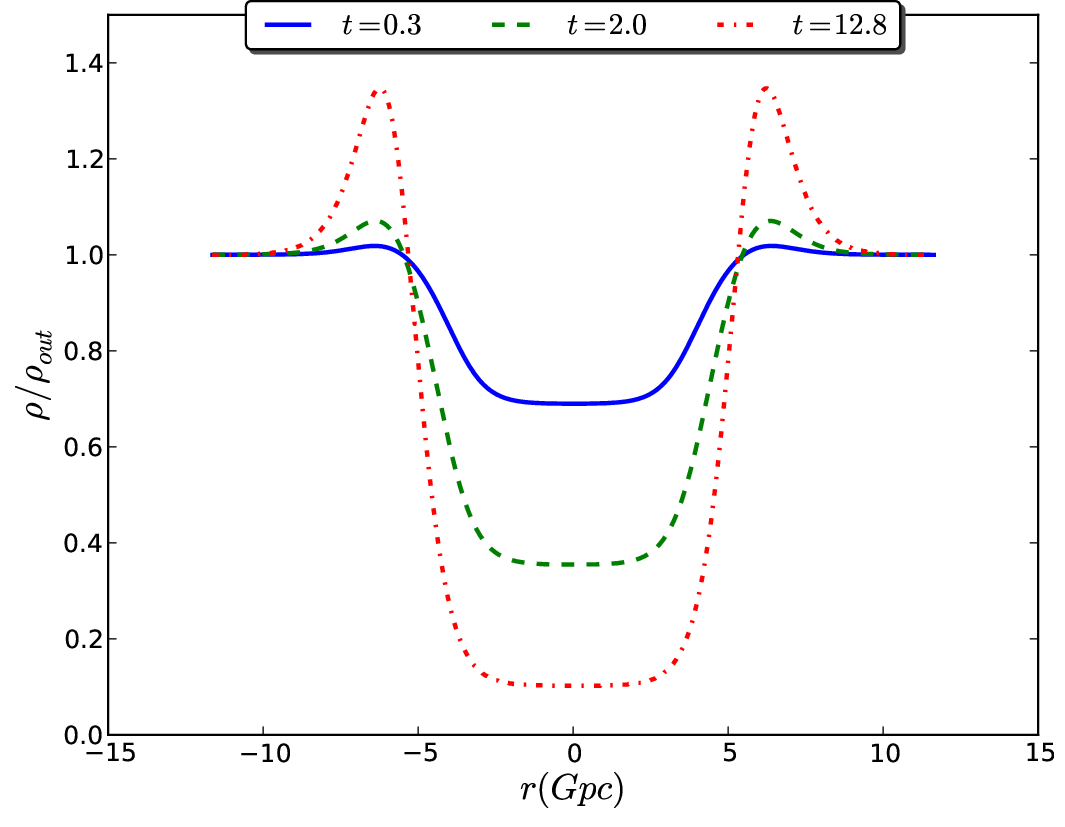}
    \caption{Density profiles at different cosmic times.  Time is in unit of
    Gyrs, and the densities are normalized to the background density.
    \label{fig.tslice-rho}}
\end{figure}

\begin{figure}
    \includegraphics[width=0.48\textwidth]{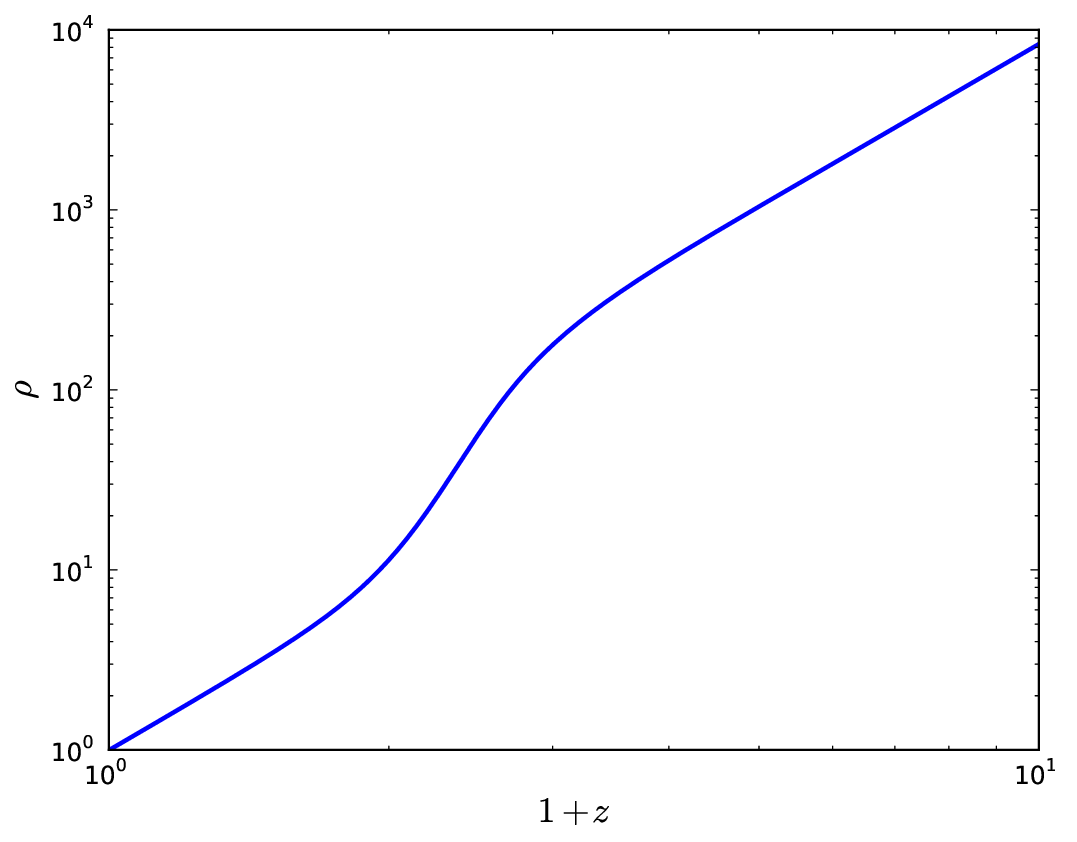}
    \caption{Densities on the past light cone, normalized to the value at the
    void center.
    \label{fig.ltcone-rho}}
\end{figure}

\begin{figure}
    \includegraphics[width=0.48\textwidth]{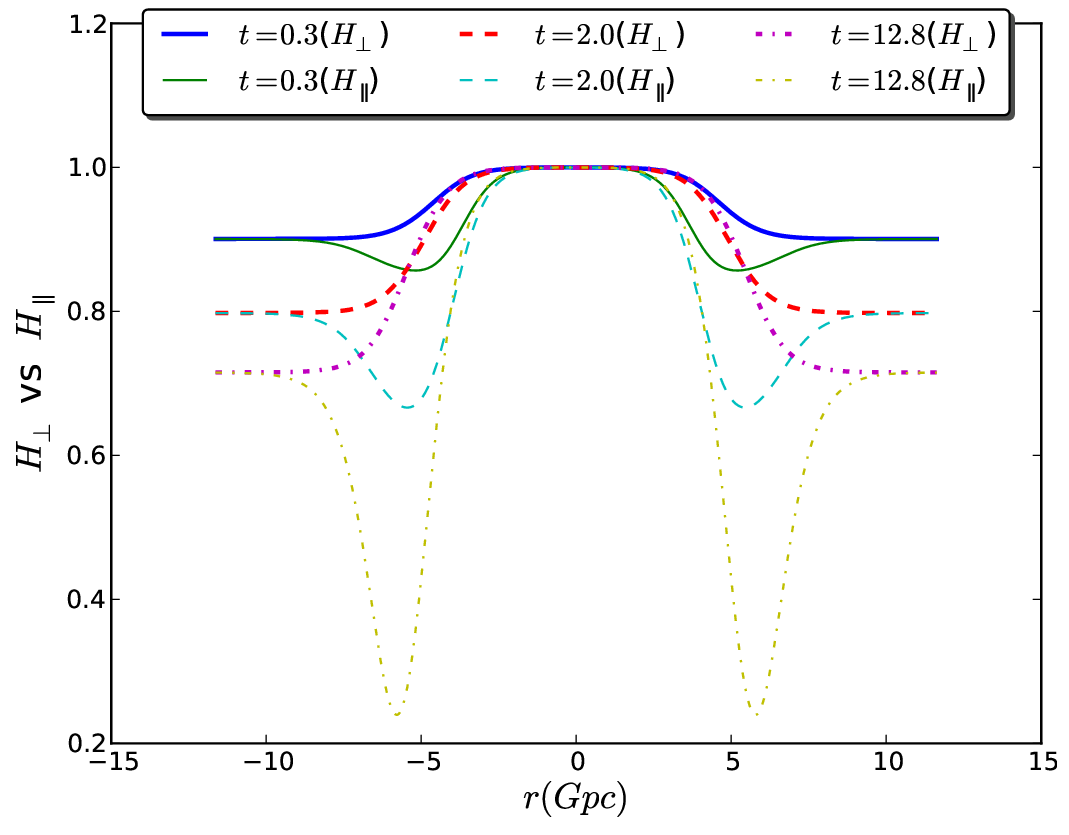}
    \caption{Comparison of the $H_\bot$ and $H_\parallel$ profiles at different
    cosmic times.  Time is in unit of Gyr, and the expansion parameters are
    normalized to the value at the void center.
    \label{fig.tslice-hpp}}
\end{figure}

\begin{figure}
    \includegraphics[width=0.48\textwidth]{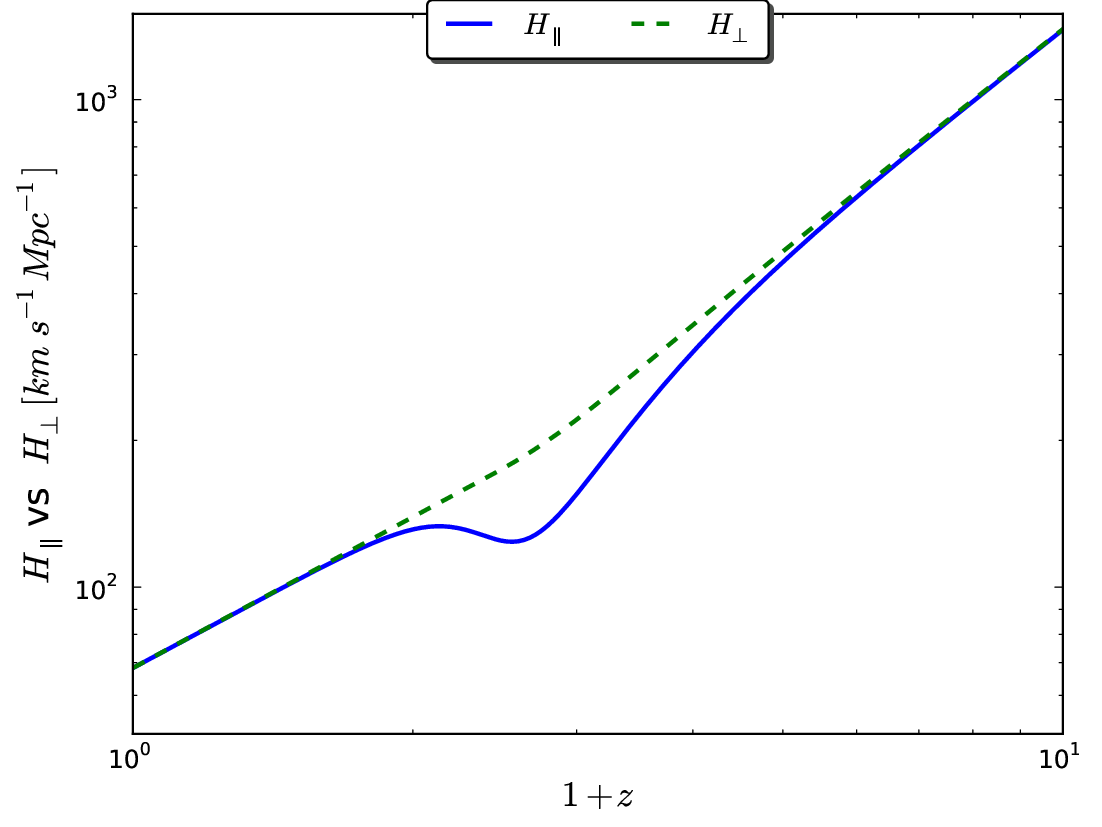}
    \caption{Comparison of the $H_\bot$ and $H_\parallel$ at different
    redshifts (on the past light cone).
    \label{fig.ltcone-hpp}}
\end{figure}

To illustrate how the universe and its evolution in CGBH model look
like, we choose $\Omega_0 = 0.05$, $H_0 = 73\ \mathrm{km\ s}^{-1} \
\mathrm{Mpc}^{-1}$, $r_0 = 6\ \mathrm{Gpc}$, $\Delta{r} = 0.1\ r_0$, and plot
in Fig.~\ref{fig.tslice-rho} and Fig.~\ref{fig.ltcone-rho} the density profile
on different cosmic time (in $Gyr$) slices and on the light cone, respectively,
and in Fig.~\ref{fig.tslice-hpp} and Fig.~\ref{fig.ltcone-hpp} the profiles of
the two expansion parameters on time slices and the light cone, respectively.

It is seen from Fig.~\ref{fig.tslice-rho} that the void gets deeper
and its shell gets denser as the universe evolves. On the other hand,
in a homogeneous universe one always has $H_\parallel = H_\bot$.
Observation of a difference, like that shown in Fig.~\ref{fig.ltcone-hpp},
within the redshift ranges of about 1-6, would imply spatial inhomogeneity.

\section{Observational Hubble parameter data}
There are four main methods to measure the Hubble parameter $H(z)$: by
measuring the differential age of passively evolving galaxies (differential age
method)\cite{2002ApJ...573...37J,2005PhRvD..71l3001S,2010JCAP...02..008S,2012JCAP...07..053M,2012JCAP...08..006M,2012arXiv1207.4541Z}, the baryon acoustic oscillation (BAO) along the line-of-sight
direction from the spectroscopic galaxy sample \cite{2009MNRAS.399.1663G},
the dipole of the luminosity distance $d_L$ of gravitational wave sources
(luminosity dipole method of standard sirens)
\cite{2006PhRvL..96s1302B,2011PhRvD..83h4045N}, and by measuring the Sandage-Loeb
signal of the Lyman-$\alpha$ forest of QSOs (Sandage-Loeb signal or redshift drift method)
\cite{2013arXiv1311.1583Y}.  The radial BAO size method
depends on the detailed evolution of perturbations not well understood in the
LTB cosmology, although progresses have been made
\cite{2008PhRvD..78d3504Z,2009JCAP...06..025C}.  The luminosity dipole method
of standard sirens by now has produced no observational data yet.  Therefore,
the OHD used in this work refer exclusively to that by the differential age
method.

The Hubble parameter for FRW models with scale factor $a$ reads
\begin{equation}
    \label{Eq.hp} H \equiv \frac{\dot a}{a} = -\frac{1}{1 + z}
    \frac{dz}{dt_{ct}},
\end{equation}
where $dt_{ct}$ is the variation of the cosmic time due to a small change in
the redshift $dz$.  For any galaxy one has $T_{CA}(z) = T_{F} + T_{GA}(z)$,
which simply states that the cosmic age $T_{CA}$ at redshift $z$ equates the
summation of the formation time of the galaxy, $T_{F}$, and the age of this
galaxy, $T_{GA}$ that can be determined spectroscopically.  If we could find a
group of galaxies that share a uniform formation time, i.~e.\ $T_{F} =
\text{const.}$, we would then get a handle of $dt_{ct}$ by simply measuring the
age difference of those galaxies:
$\label{Eq.dtfrw} dt_{ct}(z) = dT_{CA} = dT_{F} + dT_{GA}(z) = dT_{GA}(z)$.
The passively evolving galaxies can be identified by figuring out at every
redshift the oldest galaxies, which together define the `red envelop'.  One
assumes in this process the oldest galaxies formed at the same time (standard
cosmic chronometers), which is a natural assumption in an FRW universe (one may
call it the galaxy-formation version of the cosmic Copernican Principle).

Of the two expansion rates defined in Eq.\
(\ref{Eq.Hperp}), the longitudinal expansion rate $H_{\parallel}$ turns out to
have the
same form as Eq.\ (\ref{Eq.hp}) $\label{Eq. HzLTB} H_{\parallel} = -[1 / (1 +
z)] (dz / dt_{ct})$, and hence corresponds to the observed $H(z)$
\cite{2012ApJ...748..111W}. For a general LTB model, the bang time function $t_B(r)$ should not be zero,
hence the age of the Universe is: $T=t-t_B(r)$ so we have $dT/dz=(dt/dz)-(d t_B/dr)(dr/dz)$.
But as emphasized in our previous paper \cite{2012ApJ...748..111W}, the gradients
($d t_B/dr$) in the bang time, $t_B$, correspond to a currently non-vanishing decaying mode
(also see: \cite{1977A&A....59...53S,2008PhRvD..78d3504Z}. This would imply a very inhomogeneous early universe, and hence violate inflation. Further, it is true that the gradient $dt_B /dr$ is associated with density decaying modes of linear dust perturbations \cite{1977A&A....59...53S} that can be
generalised to fully non--linear LTB models \cite{2008PhRvD..78d3504Z}, though this has been updated
by recent study of dust density modes in LTB models\cite{2013CQGra..30w5001S}.
More importantly for the current work, non-zero gradient will lead to great inhomogeneities
in the galaxy formation time and make the OHD data set invalid. Therefore,
$t_B$=constant must be set, i.e., $d t_B/dr=0$, and we set it to be zero in this paper.
Of course, a mathematically zero of big bang time gradient can cripples the dynamical
freedom of the models and is not strictly necessary to prevent the violation of inflation.
Near homogeneous conditions prevailing in the last scattering surface $z\sim 1100$ can be realised by LTB
models in which the density decaying mode is not zero but becomes subdominant at
such redshift, which is argued in \cite{2011JCAP...09..011B,2012PhRvD..85b4002B,2014CQGra..31g5021S} and also show that the extreme inhomogeneity and violation of inflation is no longer valid for times close to the big bang, well before the last scattering surface. We also notice that empiric calculations are not affected if the models allow for a small gradient of the big bang time function (i.e. a small position
dependent variation in cosmic ages of different observers). Thus, it evidently does not need  the LTB models (or any other late universe inhomogeneous model) to be valid all the way back to the big bang.

%\footnote{For a general LTB model, }

The problem of using OHD in LTB models is that the basic assumption that the
oldest galaxies share a same formation time might not hold any more, as
discussed recently in Ref.~\cite{2012ApJ...748..111W}, because the background
in LTB models has considerable inhomogeneities.  However, we argue that OHD is
still valid in our context (see {\em Discussion}).  In the
following we will use the latest 23 data entries as listed in
Refs.~\cite{2012JCAP...07..053M,2012arXiv1207.4541Z},
where the data sample is larger than that of 11 OHD
used in our previous paper\cite{2012ApJ...748..111W}.

\section{Constraints on the void model}
We perform the constraints on the void model by both OHD and the background
inhomogeneity-induced kSZ (BIkSZ)
effect\cite{2011PhRvL.107d1301Z,2010MNRAS.407L..36Z}.
The direction dependent, therefore observable, temperature shift reads
\begin{equation}
    \Delta T_{\rm BI} = T_{\rm CMB} \times \int_0^{z_e} \delta_e(\hat{n}, z)
    \frac{\vec{v}_H(\hat{n}, z) \cdot \hat{n}}{c} d\tau_e,
\end{equation}
where $T_{\rm CMB} = 2.73\ {\rm K}$, $z_e = 100$ (the result essentially does
not change as long as $z(r_0) \ll z_e$), and
\begin{equation}
    v_H(z) \approx [H_\parallel(r(z), t(z)) - H_\parallel(r(z_e), t(z))]
    A(r(z),t(z))
\end{equation}
Here,$V_H$ in void model is generally contributed from both Doppler and Sachs-Wolfe
anisotropies induced by the void, and it is qualitatively dependent on the size $r_0$ of
the void considered(\cite{2011PhRvL.107d1301Z}, and therein). Similarly,
the $\beta$ function is employed instead of $V_H$ in works \cite{2012PhRvD..85b4002B,2011PhRvD..83j3515M}.
However, for small voids considered in this paper, such as Gpc-size void with size of less than a few Gpc,
Sachs-Wolfe anisotropies could be neglected, and only Doppler contribution is left as Eq.(6) above.
So only for cases of larger void cases, the general expression of $V_H$ contributed from both Doppler
and Sochs-Wolfe anisotropies should be adopted. For a more general non-spherical pattern of
inhomogeneity such as Szekeres models, the full integral $\beta$ function should be contributed from both Doppler and Sochs-Wolfe anisotropies when used to examine kSZ effect.

\begin{figure}
    \includegraphics[width=0.48\textwidth]{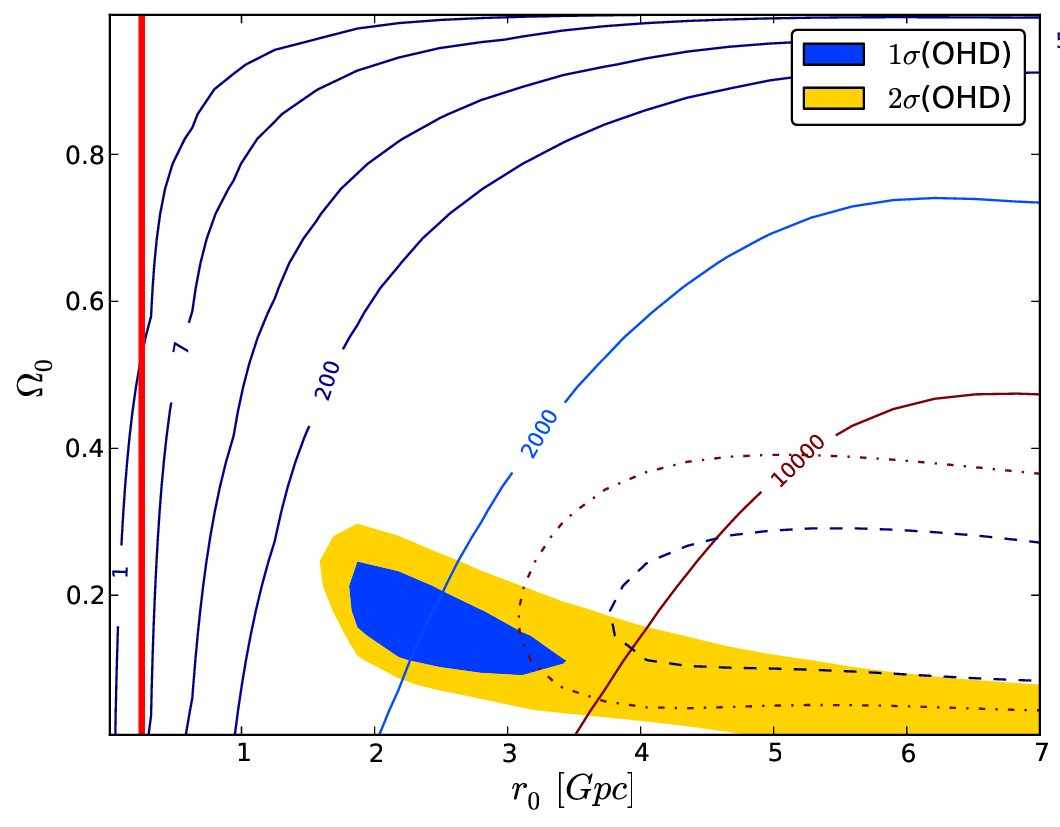}
    \caption{Background inhomogeneity-induced kSZ effect $\Delta T_{\rm BI}^2$
    in $\mu {\rm K}^2$.  The line marked with $7$ corresponds to the 95\%
    observational upper limit of the South Pole Telescope
    \cite{2010ApJ...718..632H,2011ApJ...736...61S} which is significant to the
    constraint of void models \cite{2011PhRvL.107d1301Z}.  Also plotted are the
    1$\sigma$ and 2$\sigma$ confidence regions from the OHD as well as those
    from the supernovae Union2 dataset.  The vertical (red) bar marks the
    characteristic void radius $r = 250\ \mathrm{Mpc}$ in the small-void models
    of Refs.~\cite{2009JCAP...09..025A,2011PhLB..697..265B} that are comparable
    with recent observational data of near-infrared galaxy luminosity function
    \cite{2012arXiv1207.1588K}.
    \label{fig.ksz}}
\end{figure}

We calculate BIkSZ power spectrum $\Delta T_{\rm BI}^2$ at $l = 3000$ and its
constraints on the $(r_0, \Omega_0)$
parameter plane, with $\Delta r / r_0 = 0.21$, $H_0 = 74\ \mathrm{km\ s}^{-1} \
\mathrm{Mpc}^{-1}$ fixed at their respective best-fit values, which are in turn
obtained from the Hubble parameter dataset. The resulting contours, as well as
confidence regions from the OHD and the supernovae Union2 dataset, are plotted
in Fig.~\ref{fig.ksz}. The OHD data favor
a smaller (and more tightly constrained) void than what the supernovae Union2
data do. Indeed, there is a clear discrepancy between those two datasets, as
found in Ref.~\cite{2012ApJ...748..111W}.  This is a sign of inadequacy for
this specific LTB model.  Also, as pointed out in
Ref.~\cite{2011PhRvL.107d1301Z}, one can see from Fig.~\ref{fig.ksz} that the
Gpc-sized voids, as those favored by the supernovae data, are incompatible with
the BIkSZ measurement, hence are largely excluded.  Now we can tell from the
figure that the OHD dataset give slightly weaker, though basically the same,
conclusion.  Furthermore, the observed normalization of the near-IR galaxy
luminosity function indicates that a void, if exists, amounts to a few hundred
Mpcs\cite{2012arXiv1207.1588K}.  This could in principle be consistent with the
BIkSZ measurement.

\section{Future BAO constraint}
As is shown in Fig.~\ref{fig.ltcone-hpp} above, unlike the homogeneous cosmological models,
$H_\parallel$ can differ from $H_\bot$in LTB models at redshift $z$ roughly ranging from 1 to 6.
Therefore one straightforward way is to define the ratio between these two
expansion rates $\mathcal{E} = H_\parallel / H_\bot$,
which always equals to $1$ for the homogeneous cosmological models, but
deviates from $1$ in LTB models. Specifically, we can further
write $\mathcal{E}$ as
\begin{equation}\label{eqn:Elb}
    \mathcal{E} = H_\parallel / H_\bot = 1 + \frac{A}{A'} \frac{H'_\bot}
    {H_\bot},
\end{equation}
since $H_\parallel= \frac{\dot{A}'}{A'} = H_\bot + \frac{A}{A'} H'_\bot$.

Therefore, the violation of $\mathcal{E} = 1$  could also be an indicator
of LTB-type models.
To adopt this criteria however, one needs independent measurements of
$H_{\parallel}$ and $H_\bot$ at the same redshift or just the variation of $
H_\bot$ on the light cone.  The BAO feature imprinted in the non-relativistic
matter such as galaxies distribution yields a further geometric test of
homogeneity, and future large-volume BAO surveys
will also allow us detect the BAO scale in both radial and transverse
directions, although the transverse $H_\bot$ could not be measured
in current measurement accuracy of BAO. Thus, we expect future BAO measurement would supply the
information of the criteria $\mathcal{E}$ by radial Hubble parameter
$H_\parallel$
and transverse Hubble parameter $H_\bot$, and improve greatly the testing of
violation of homogeneity.

\section{Discussion}
In our calculation of OHD constraints on the void model, we assume that OHD
could be used in LTB models.  Actually, the same formation time of the oldest
galaxies is a basic assumption in obtaining OHD.  However, in LTB models where
the universe has a considerable background inhomogeneity, this assumption
becomes unreasonable and some arguments are also recently given in
Ref.~\cite{2012arXiv1208.4534D} where galaxy ages are used.  First, despite the
overall uniform-formation-age assumption, the validity of an $H(z)$ data point
requires the same formation time only inside the redshift bin where OHD is
locally defined and obtained (Eq.~[\ref{Eq.hp}]), even if the global density --
hence the formation time of the oldest galaxies -- at different redshifts
varies much.  Secondly, a standard viewpoint (referred to as the onion
approximation) is to treat the LTB void universe as a group of thin shells
structured together, and inside each of these spherical shells the matter is
homogeneously distributed \cite{2007JCAP...12..017B}.  For the OHD used in this
paper, the size of each redshift bin is between $0.1$ and $0.15$, where the
first limit is so chosen that the age evolution between the two bins is larger
than the error in the age determination \citep{2005PhRvD..71l3001S}.  As the
precision of the age determination improves, we expect an even smaller bin
size.  To be sure about the validity of OHD used in LTB models, one
needs the exact knowledge about the thickness of the shell given the size of a
redshift bin, as well as the steepness of the density profile at the time the
oldest galaxies formed. We ever discussed this issue in
Ref.~\cite{2012ApJ...748..111W}.

On the other hand, future observation is expected to yield $\sim{}2000$
measurements for passively evolving galaxies in the redshift range $0 < z <
1.5$ in the future \cite{2005PhRvD..71l3001S}.  It has been estimated that
about 1000 OHD entries at a $15\%$ accuracy level will be determined with
$10\%$ error of the galaxies ages.  In Ref.~\cite{2011ApJ...730...74M} the
power of OHD in the context of ${\rm \Lambda CDM}$ model has been assessed.
With the increase of high quantity OHD, the power of OHD constraining void
model should also be greatly improved for constraining the void models
\cite{2012ApJ...748..111W}.

Although the large void model appear to be ruled out by some cosmological
observations, future OHD measurement in both radial and transverse directions,
as an alternative and complementary cosmological test, could give a tight
constraint on LTB model with a small void.  If the transverse BAO information
can be realized from future large-scale structure observations, we should be
able to arrive at a definite test of spatial homogeneity of the Universe.  In
this context, the role played by the transverse BAO is complementary to the
radial BAO discussed in Ref.~\cite{2011JCAP...09..035H}.

\begin{acknowledgments}
We sincerely thank the two PRL and last anonymous referees whose suggestions 
and objective, judicial 
assessment greatly helped us improve our manuscript. Tong-Jie Zhang thank Prof. Martin White for his hospitality during visiting
Departments of Physics and Astronomy, University of California, Berkeley and Lawrence
Berkeley National Laboratory. This work was supported by the National
Science Foundation of China (Grants No.\ 11173006), the Ministry of Science
and Technology National Basic Science program (project 973) under
grant No.\ 2012CB821804.
\end{acknowledgments}

\appendix
\section{$H_{\parallel}$ and $H_\bot$  in a coordinate independent manner}

The ``longitudinal'' and ``transverse'' Hubble parameter in Eq.(2) has only a covariant
meaning for spherical symmetry: they are the components, tangent and orthogonal
to the orbits of SO(3), of the expansion tensor $H^a_b = H\,h^a_b +
\sigma^a_b$,  where $H=(1/3)\nabla_a u^a$ is the Hubble expansion scalar and
$\sigma^a_b$ is the shear tensor. Since $H^r_r=H_{||}$  and
$H^\theta_\theta=H^\phi_\phi = H_\perp$, and we can always choose an orthonormal
tetrad with $u^a$ as the timelike tetrad vector, one vector $n^a$ orthogonal to
the orbits and two vectors $m_{(1)}^a,\,m_{(2)}^a$ tangent to them, then
$H_{||}=H_{ab} n^a n^b$  and $H_\perp=H_{ab} m_{(1)}^a m_{(1)}^b=H_{ab}
m_{(2)}^a m_{(2)}^b$, which renders the parameter ${\cal E}$ introduced in Eq.(7)
as the ratio

$$ {\cal E} = \frac{H_{ab} n^a n^b}{H_{ab} m_{(1)}^a m_{(1)}^b } =
\frac{H_{ab} n^a n^b}{H_{ab} m_{(2)}^a m_{(2)}^b }$$

For non-spherical models, these quantities can always be computed in terms of
an orthonormal tetrad, but their  interpretation as ``longitudinal'' and
``transverse'' becomes coordinate dependent. Another possible comparison that
provides a measure of local inhomogeneity is given by the ratios $\sigma_1/H$
and $\sigma_2/H$, where  $\sigma_1$ and $\sigma_2$ are the eigenvalues of the
shear tensor (being trace-free and it admits two eigenvalues in general). For LTB
models, $\sigma_1=\sigma_2=\sigma = -(1/3)(H_{||}-H_\perp)$ is the unique
eigenvalue: $\sigma^a_b = \sigma e^a_b$ with $e^a_b=h^a_b-3n^an_b$ and
$H=2H_{||}+H_\perp$.  For Szekeres models, there is also a single shear eigenvalue,
so by replacing  ${\cal E}$ with
$\sigma/H$ as a measure of inhomogeneity one can readily generalise the
interpretation of observational tests from LTB to Szekeres models. In fact, all of LTB quantities given in coordinate independent manner can be readily generalised to that in Szekeres models\cite{2012CQGra..29f5018S,2012JCAP...12..001W}.

%\end{widetext}

\bibliography{ms_PRL}

\end{document}